\documentclass[preprint,epsfig,onecolumn,floats,showpacs]{revtex4}

\begin{document}
\title{Partial transposition on bi-partite system}
\author{Y.-J. Han, X. J. Ren, Y. C. Wu, G.-C. Guo}
\address{Key Laboratory of Quantum Information, University of Science and\\
Technology of China, Hefei 230026, China}

\begin{abstract}
Many of the properties of the partial transposition are not clear so far.
Here the number of the negative eigenvalues of $\rho ^{T}$ is considered
carefully when $\rho $ is a two-partite state. There are strong evidences to
show that the number of negative eigenvalues of $\rho ^{T}$ is $\frac{N(N-1)%
}{2}$ at most when $\rho $ is a state in Hilbert space $C^{N}\otimes C^{N}$.
For the special case, $2\times 2$ system, we use this result to give a
partial proof of the conjecture $\left\vert \rho ^{T}\right\vert ^{T}\geq 0$%
. We find that this conjecture is strongly connected with the
entanglement of the state corresponding to the negative eigenvalue
of $\rho ^{T}$ or the negative entropy of $\rho $.
\pacs{03.67.Mn,03.65.Ud, 03.67.-a}

\end{abstract}
\maketitle

\section{Introduction}

Entanglement is one of the most interesting properties of many-body systems.
It has many powerful applications, such as quantum communication\cite%
{neilsenbook} and quantum computation\cite{shor,grov}. So it is the
cornerstone of quantum information technology\cite{1,2,3}. Although it is
very important, unfortunately, it is very difficult to investigate the
measurement of entanglement. Sometimes, it is even hard to decide whether a
state is entangled or not. Although some useful methods (such as
entanglement witness\cite{witt,witt1}, partial transposition \cite%
{ppt,ppt1,ppt2}) have been introduced to this problem, it is far from
completing this problem even for two-body systems. Partial transposition
(PT) is the most powerful tool to detect the entanglement of a state\cite%
{cirac} , especially for bipartite systems. The famous Peres-Horodecki
criterion \cite{ppt,ppt1} guarantees that the positive partial transposition
(PPT) condition is sufficient and necessary to decide whether a state in
Hilbert space $C^{2}\otimes C^{2}$ or $C^{2}\otimes C^{3}$ is entangled or
not. But PPT condition is only sufficient (not necessary) for higher
dimensional bipartite systems.

There are two reasons for us to consider the PT transformation more
carefully in quantum information. One is that we want to know whether the PT
method can be generalized and to find the similar method to the higher
dimensional Hilbert space, since there are many advantages of this method
(such as the convenience of computation). In other words, we want to know
the reason why we can or can not extend this method and how if we can. To
make these ideas possible, we must carefully investigate the physical
significance and mathematical characters of PT transformation. The other
reason is that we know little about the PT transformation even in $2\times 2$
system, since there is no explicitly known mathematical operator
corresponding to PT transformation. Many characters (mathematical or
physical) of PT are still unclear so far. A simple question of $2\times 2$
system is introduced by Audenaert {\it et al} \cite{vollbart,someproblem}:
Prove that
\begin{equation}
\left| \rho ^T\right| ^T\geq 0  \eqnum{1}
\end{equation}
for any two-qubit state $\rho $, where $T$ is PT and $\left| .\right| $ is
the operator absolute value. This question is clear, but the proof is not
found yet. We do not know whether equation(1) is correct or not. In this
paper, we devote to a even more simple problem about PT transformation at
first: let $\rho $ be a bi-partite state in Hilbert space $C^N\otimes C^M$,
how many negative eigenvalues are there in $\rho ^T$ at most? We find that
there are strong evidences showing that the number of the negative
eigenvalues of $\rho ^T$ is $\frac{N(N-1)}2$ at most when $N=M$. Then we
give a partial proof for the problem introduced by Audenaert {\it et al}
under the assumption that the number of negative eigenvalue of $\rho ^T$
(where $\rho $ is a two-qubit state) is 1 at most. At last,we will give our
conclusion.

\section{The number of negative eigenvalues of $\protect\rho ^T$}

It is well known that the eigenvalues of $\rho ^T$ ($\rho $ is a general
state in bi-partite system) can be used to detect the entangled property of
bi-partite state. When $\rho $ is a separable state, the eigenvalues of $%
\rho ^T$ are positive. But for some entangled states, the eigenvalues of $%
\rho ^T$ may be negative. How many negative eigenvalues can be in $\rho ^T$
for different systems? Does the number of the negative eigenvalues have any
connection with the entanglement property of $\rho $ ? These questions are
very clear and simple, but their answers are very difficult.

For convenience, we first introduce a useful lemma in matrix analysis\cite%
{horn} before the investigation of these problems. This lemma is about the
eigenvalues relations between a Hermitian matrix and its principal submatrix.

{\it Lemma: }Let $A$ be a $n\times n$ Hermitian matrix, let $r$ be a integer
with $1\leq r\leq n$, and let $A_r$ be a $r\times r$ principal submatrix of $%
A$. Then for each integer $k$ such that $1\leq k\leq n$ we have
\begin{equation}
\lambda _k(A)\leq \lambda _k(A_r)\leq \lambda _{k+n-r}(A),  \eqnum{2}
\end{equation}
where $\lambda _k(A)$ and $\lambda _k(A_r)$ are the $k$th eigenvalue of $A$
and $A_r$ in increasing order, respectively.

Given this Lemma, we can get the following theory about the number of the
negative eigenvalues of $\rho ^T$ immediately:

{\it Theorem 1}: Let $\rho $ be a bi-partite state in Hilbert space $%
C^M\otimes C^N$, then the number of the negative eigenvalues of $\rho ^T$ is
less than $MN-\max (N,M)$.

{\it Proof: }Without loss of generality, let $N\geq M$. Suppose the PT
operates on the first particle, then using the definition of PT, the $M$
diagonal $N\times N$ blocks of $\rho ^T$ (these blocks are $\left\langle
1_A\right| \rho ^T\left| 1_A\right\rangle $, $\left\langle 2_A\right| \rho
^T\left| 2_A\right\rangle $, $\cdots $, $\left\langle M_A\right| \rho
^T\left| M_A\right\rangle $) are the same as the corresponding blocks of $%
\rho $ (these blocks are $\left\langle 1_A\right| \rho \left|
1_A\right\rangle $, $\left\langle 2_A\right| \rho \left| 2_A\right\rangle $,
$\cdots $, $\left\langle M_A\right| \rho \left| M_A\right\rangle $). Since $%
\rho $ is positive, all these diagonal $N\times N$ blocks are positive.

Let one of the $N\times N$ blocks (such as $\left\langle 2_A\right| \rho
^T\left| 2_A\right\rangle $) be $A_{N\times N}$ and its eigenvalues,
arranged in increasing order, be \{$\lambda _1,\lambda _2,\cdots ,\lambda _N$%
\}. In addition, let the eigenvalues of $\rho ^T$, also arranged in
increasing order, be \{$\widehat{\lambda }_1,\widehat{\lambda }_2,\cdots ,%
\widehat{\lambda }_{N\times M}$\}. It is clear that the block $A_{N\times N}$
is a principal submatrix of $\rho ^T$ (obtained by deleting some $(M-1)N$
rows and the corresponding columns from $\rho ^T$). By the lemma, we can get
\begin{equation}
\widehat{\lambda }_k\leq \lambda _k\leq \widehat{\lambda }_{k+(M-1)N}
\eqnum{3}
\end{equation}
for each integer $k$ such that $1\leq k\leq N$. Specially, let $k=1$, we can
get
\begin{equation}
0\leq \lambda _1\leq \widehat{\lambda }_{1+(M-1)N}.  \eqnum{4}
\end{equation}
That is, $\widehat{\lambda }_{1+(M-1)N}\geq 0$. So the number of the
negative eigenvalues of $\rho ^T$ is $(M-1)N$ at most.

Q. E. D

Is this the exact limit for the number of the negative eigenvalues of $\rho
^T$? When we investigate this problem for $2\times 2$ system more carefully,
we find that it is not the case. We consider the following cases in the
two-qubit system to get some concrete idea .

1) $\rho $ is a pure state. It is very clear that $\rho ^T$ has $0$ negative
eigenvalues when $\rho $ is separable and $\rho ^T$ has $1$ negative
eigenvalues when $\rho $ is entangled.

2) $\rho $ is a Bell diagonalized state. So $\rho $ and $\rho ^T$ have the
forms of
\begin{equation}
\left[
\begin{array}{llll}
\alpha _1 & 0 & 0 & \beta _1 \\
0 & \alpha _2 & \beta _2 & 0 \\
0 & \beta _2^{*} & \alpha _3 & 0 \\
\beta _1^{*} & 0 & 0 & \alpha _4%
\end{array}
\right] \text{ and }\left[
\begin{array}{llll}
\alpha _1 & 0 & 0 & \beta _2^{*} \\
0 & \alpha _2 & \beta _1^{*} & 0 \\
0 & \beta _1 & \alpha _3 & 0 \\
\beta _2 & 0 & 0 & \alpha _4%
\end{array}
\right] ,  \eqnum{5}
\end{equation}
respectively. We can get the relations $\alpha _1\alpha _4-\left| \beta
_1\right| ^2\geq 0$ and $\alpha _2\alpha _3-\left| \beta _2\right| ^2\geq 0$
by the positive semidefinite of $\rho $. So either $\alpha _1\alpha
_4-\left| \beta _2\right| ^2\geq 0$ or $\alpha _2\alpha _3-\left| \beta
_1\right| ^2\geq 0$. That is, there is a $3\times 3$ principal submatrix are
positive semidefinite in $\rho ^T$. So the number of the negative eigenvalue
of $\rho ^T$ is 1 at most by the similar reason of equation(2).

In fact, to extensively investigate this problem, we use the Monte Carlo
method\cite{land} to choose one million random samples and calculate their
negative eigenvalues. We find that the number of negative eigenvalues of $%
\rho ^T$ is $1$ at most in $2\times 2$ system. More generally, we use Monte
Carlo method to consider the system $N\times M$ and we get the maximal
number of negative eigenvalues of $\rho ^T$ in the following table when $N$
and $M$ are small numbers.
\[
\begin{tabular}{c|cccccccccc}
M%
\mbox{$\backslash$}
N & 2 & 3 & 4 & 5 & \multicolumn{1}{c}{6} & 7 & 8 & 9 & 10 & $\cdots $ \\
\hline
2 & 1 & 2 & 3 & 3 & \multicolumn{1}{c}{3} & 4 & 4 & 4 & 5 &  \\
3 &  & 3 & 4 & 4 & \multicolumn{1}{c}{5} & 5 & 6 & 6 & 7 &  \\
4 &  &  & 6 & 6 & \multicolumn{1}{c}{7} & 8 & 8 & 8 & 9 &  \\
5 &  &  &  & 10 & 10 & 10 & 11 & 11 & 11 &  \\
6 &  &  &  &  & \multicolumn{1}{c}{15} & 15 & 15 & 15 & 16 &  \\
7 &  &  &  &  & \multicolumn{1}{c}{} & 21 & 21 & 21 & 21 &  \\
8 &  &  &  &  & \multicolumn{1}{c}{} &  & 28 & 28 & 28 &  \\
9 &  &  &  &  & \multicolumn{1}{c}{} &  &  & 36 & 36 &  \\
10 &  &  &  &  & \multicolumn{1}{c}{} &  &  &  & 45 &  \\
$\vdots $ &  &  &  &  & \multicolumn{1}{c}{} &  &  &  &  & $\ddots $%
\end{tabular}
\]

\[
\text{Table }
\]

With this table, we have good reasons to give the following conjecture.

{\it Conjecture}: The number of the negative eigenvalues of $\rho ^T$ is at
most $\frac{N(N-1)}2$ when $\rho $ is a state in Hilbert space $C^N\otimes
C^N$.

Since PT transformation has no direct geometrical or algebraic meaning, it
is difficult to use the powerful geometrical or algebraic tools to
completely prove this result even for the simplest case of $2\times 2$
system. But for this special two-qubit case, we can give some partial
results. It is well known that the character of $\rho ^T$ is due to the
entanglement of $\rho $ in two-qubit case, so the number of the negative
eigenvalues of $\rho ^T$ is independent of the local unitary. If we use $%
\left| 0\right\rangle _1\left| 0\right\rangle _2,\left| 0\right\rangle
_1\left| 1\right\rangle _2,\left| 1\right\rangle _1\left| 0\right\rangle _2$
and $\left| 1\right\rangle _1\left| 1\right\rangle _2$ [where $\left|
0\right\rangle _1$ and $\left| 1\right\rangle _1$ are the eigenvectors of $%
\rho _1=tr_2(\rho )$, $\left| 0\right\rangle _2$ and $\left| 1\right\rangle
_2$ are the eigenvectors of $\rho _2=tr_1(\rho )$] as the basis, and adjust
their phase carefully, it is easy to proof that any density matrix $\rho $
of $2\times 2$ system can be transformed in the following form
\begin{equation}
\rho =\left[
\begin{array}{llll}
a_{11} & A & B & \alpha \\
A & a_{22} & \beta & -B \\
B & \beta ^{*} & a_{33} & -A \\
\alpha ^{*} & -B & -A & a_{44}%
\end{array}
\right] ,  \eqnum{6}
\end{equation}
where $A$, $B$, $a_{11}$, $a_{22}$, $a_{33}$ and $a_{44}$ are real numbers.
For this density matrix $\rho $,we have the following result:

{\it Theorem 2}: Let $\rho $ be the form (6), if $AB=0$ or $%
\mathop{\rm Re}%
(\alpha )=%
\mathop{\rm Re}%
(\beta )$, then $\rho ^T$ has $1$ negative eigenvalue at most.

Many important states in quantum information are included in this theorem,
such as i)Pure states, ii)Bell diagonalized states, iii) Werner states.
Though this theorem is not the complete proof of the conjecture of $2\times
2 $ system, it includes enough cases for our use. The proof of this theorem
is rather simple.

{\it Proof}: Since the form of $\rho $ is given as equation(6), then we can
get
\begin{equation}
\rho ^T=\left[
\begin{array}{llll}
a_{11} & A & B & \beta ^{*} \\
A & a_{22} & \alpha ^{*} & -B \\
B & \alpha & a_{33} & -A \\
\beta & -B & -A & a_{44}%
\end{array}
\right] ,  \eqnum{7}
\end{equation}
where the PT transformation operates on the first qubit. For any $3\times 3$
principal submatrix $A^T$ of $\rho ^T$ ,we have the following interlacing
relations\cite{horn} between the eigenvalues of $A^T$ and $\rho ^T$ . Let $%
\lambda _i(i=1,2,3)$ and $\widehat{\lambda }_i(i=1,2,3,4)$ be the
eigenvalues of $A^T$ and $\rho ^T$, respectively, and assume that they are
arranged in increasing order. Then
\begin{equation}
\widehat{\lambda }_1\leq \lambda _1\leq \cdots \leq \lambda _3\leq \widehat{%
\lambda }_4.  \eqnum{8}
\end{equation}
So we can conclude that $\rho ^T$ has at most $1$ negative eigenvalues if
some $3\times 3$ principal submatrix of $\rho ^T$ is positive semidefinite.

Let us consider the $3\times 3$ principal submatrices of $\rho ^T$ and $\rho
$. We consider two sets
\begin{equation}
A_1^T=\left[
\begin{array}{lll}
a_{11} & A & B \\
A & a_{22} & \alpha ^{*} \\
B & \alpha & a_{33}%
\end{array}
\right] ,A_1=\left[
\begin{array}{lll}
a_{11} & A & B \\
A & a_{22} & \beta \\
B & \beta ^{*} & a_{33}%
\end{array}
\right]  \eqnum{9}
\end{equation}
and
\begin{equation}
A_2^T=\left[
\begin{array}{lll}
a_{11} & A & \beta ^{*} \\
A & a_{22} & -B \\
\beta & -B & a_{44}%
\end{array}
\right] ,A_2=\left[
\begin{array}{lll}
a_{11} & A & \alpha \\
A & a_{22} & -B \\
\alpha ^{*} & -B & a_{44}%
\end{array}
\right] .  \eqnum{10}
\end{equation}
Since $\rho $ is positive, all the principal submatrix of $\rho $ are
positive. Specially, $\left[
\begin{array}{ll}
a_{11} & A \\
A & a_{22}%
\end{array}
\right] $ is positive too. Now we only need calculate the determinates of $%
A_1^T$ and $A_2^T$ under the condition that $A_1$ and $A_2$ are positive.
Simple calculations can get
\begin{equation}
\mathop{\rm Det}%
(A_1)-%
\mathop{\rm Det}%
(A_1^T)=2AB%
\mathop{\rm Re}%
(\beta -\alpha )+a_{11}(\left| \alpha ^2\right| -\left| \beta ^2\right| ),
\eqnum{11}
\end{equation}
\begin{equation}
\mathop{\rm Det}%
(A_2)-%
\mathop{\rm Det}%
(A_2^T)=2AB%
\mathop{\rm Re}%
(\beta -\alpha )+a_{11}(\left| \beta ^2\right| -\left| \alpha ^2\right| ).
\eqnum{12}
\end{equation}
Now, Using the conditions that $AB=0$ or $%
\mathop{\rm Re}%
(\alpha )=%
\mathop{\rm Re}%
(\beta )$, we can get
\begin{equation}
\mathop{\rm Det}%
(A_1)-%
\mathop{\rm Det}%
(A_1^T)=a_{11}(\left| \alpha ^2\right| -\left| \beta ^2\right| ),  \eqnum{13}
\end{equation}
\begin{equation}
\mathop{\rm Det}%
(A_2)-%
\mathop{\rm Det}%
(A_2^T)=a_{11}(\left| \beta ^2\right| -\left| \alpha ^2\right| ).  \eqnum{14}
\end{equation}
Obviously, $%
\mathop{\rm Det}%
(A_1)-%
\mathop{\rm Det}%
(A_1^T)$ or $%
\mathop{\rm Det}%
(A_2)-%
\mathop{\rm Det}%
(A_2^T)$ is no more than 0. That is, $%
\mathop{\rm Det}%
(A_1^T)\geq
\mathop{\rm Det}%
(A_1)$ or $%
\mathop{\rm Det}%
(A_2^T)\geq
\mathop{\rm Det}%
(A_2)$. So $A_1^T$ or $A_2^T$ is positive semidefinite. This is the end of
the proof.

Q. E. D.

But for the higher dimensional situations, the similar result is rather
difficult to be obtained. There are a lot of things to do to completely
prove our conjecture.

\section{Partial proof of $\left| \protect\rho ^T\right| ^T\geq 0$ in
two-qubit system}

If the number of negative eigenvalue of $\rho ^T$ ($\rho $ is in $2\times 2$
system) is really at most 1, we can give a partial proof of equation.(1).
Obviously, when the two-qubit state $\rho $ is separable, the equation(1) is
correct. But our result show that when the entanglement of the eigenvector
corresponding to the negative eigenvalue of $\rho ^T$ is sufficient high
(which connected with the condition about the negative entropy of $\rho $)
the equation is true too.

For any Hermitian matrix\cite{operator} $H$, we can make a expansion
\begin{equation}
H=H_{+}-H_{-}  \eqnum{15}
\end{equation}
where $H_{+}$ and $H_{-}$ are positive semidefinite and $%
H_{+}H_{-}=H_{-}H_{+}=0.$ So assume
\begin{equation}
\rho ^T=A-\rho _{-}  \eqnum{16}
\end{equation}
where $%
\mathop{\rm Rank}%
(A)=3$, $A\rho _{-}=0$ , $\rho _{-}=\left| \Psi \right\rangle \left\langle
\Psi \right| $ and $\left| \Psi \right\rangle =\alpha \left| 00\right\rangle
+\beta \left| 11\right\rangle $ (where $\left| \Psi \right\rangle $ is not
normalized and we have used proper local unitary transformations to make $%
\alpha $ and $\beta $ positive real number) at the present situation. Now we
can describe our result as the following

{\it Theorem 3}: Let $\rho $ be a two-qubit density matrix, its partial
transposition $\rho ^T$ have only one negative eigenvalue $E$ and the
corresponding eigenvector is $\left| \Psi \right\rangle =\alpha \left|
00\right\rangle +\beta \left| 11\right\rangle $ (where $\left| \Psi
\right\rangle $ is not normalized, $\alpha \geq \beta $ and $\alpha ^2+\beta
^2=\left| E\right| $ ). If $\alpha $ and $\beta $ satisfy the conditions:
\begin{equation}
\alpha \beta =0  \eqnum{17.1}
\end{equation}
or
\begin{equation}
1\leq \frac \alpha \beta \leq \sqrt{\sqrt{2}+1}.  \eqnum{17.2}
\end{equation}
Then $\left| \rho ^T\right| ^T\geq 0.$

{\it Proof}: Using the equation (16), we can rewrite $\rho ^T$ and $\rho $
in explicit form as follows
\begin{equation}
\rho ^T=A-\rho _{-}=\left[
\begin{array}{llll}
A_{11}-\alpha ^2 & A_{12} & A_{13} & A_{14}-\alpha \beta \\
A_{21} & A_{22} & A_{23} & A_{24} \\
A_{31} & A_{32} & A_{33} & A_{34} \\
A_{41}-\alpha \beta & A_{42} & A_{43} & A_{44}-\beta ^2%
\end{array}
\right]  \eqnum{18}
\end{equation}
and
\begin{equation}
\rho =A^T-\rho _{-}^T=\left[
\begin{array}{llll}
A_{11}-\alpha ^2 & A_{12} & A_{31} & A_{32} \\
A_{21} & A_{22} & A_{41}-\alpha \beta & A_{42} \\
A_{13} & A_{14}-\alpha \beta & A_{33} & A_{34} \\
A_{23} & A_{24} & A_{43} & A_{44}-\beta ^2%
\end{array}
\right] ,  \eqnum{19}
\end{equation}
where we have supposed that the PT transformation operates on the first
qubit. Since $\rho $ is a density matrix of a state, it is positive
semidefinite. On the other hand, from the formula (18), we can get $\left|
\rho ^T\right| =A+\rho _{-}$. So we can express $\left| \rho ^T\right| ^T$
as follows
\begin{equation}
\left| \rho ^T\right| ^T=A^T+\rho _{-}^T=\left[
\begin{array}{llll}
A_{11}+\alpha ^2 & A_{12} & A_{31} & A_{32} \\
A_{21} & A_{22} & A_{41}+\alpha \beta & A_{42} \\
A_{13} & A_{14}+\alpha \beta & A_{33} & A_{34} \\
A_{23} & A_{24} & A_{43} & A_{44}+\beta ^2%
\end{array}
\right] .  \eqnum{20}
\end{equation}
If $\alpha $ or $\beta $ is equal to zero, that is $\alpha \beta =0$, It is
clear that $\left| \rho ^T\right| ^T$ is positive semidefinite. Without loss
of generality, let $\beta =0$, then $\rho $ and $\left| \rho ^T\right| ^T$
are the same except the element $(1$,$1)$. Since $\rho $ is positive
semidefinite, the 3-by-3 principal submatrix
\begin{equation}
A_{sub}=\left[
\begin{array}{lll}
A_{22} & A_{41} & A_{42} \\
A_{14} & A_{33} & A_{34} \\
A_{24} & A_{43} & A_{44}%
\end{array}
\right]  \eqnum{21}
\end{equation}
is positive semidefinite. In order to prove the positive semidefinite of $%
\left| \rho ^T\right| ^T$, we need only consider the determinate of $\left|
\rho ^T\right| ^T$. Simple calculation can find that
\begin{equation}
\mathop{\rm Det}%
(\left| \rho ^T\right| ^T)-%
\mathop{\rm Det}%
(\rho )=2\left| \alpha ^2\right|
\mathop{\rm Det}%
(A_{sub})\geq 0.  \eqnum{22}
\end{equation}
Since $\rho $ is positive semidefinite, then $%
\mathop{\rm Det}%
(\left| \rho ^T\right| ^T)$ is positive and $\left| \rho ^T\right| ^T$ is
positive semidefinite. From now on we assume neither $\alpha $ nor $\beta $
is equal to 0.

Now we can consider the condition between $A$ and $\rho _{-}$: $A\rho _{-}=0$%
. This condition can be expressed by the elements of $A$ and $\rho _{-}$ as
\begin{eqnarray}
\alpha A_{11}+\beta A_{41} &=&0,  \eqnum{23} \\
\alpha A_{41}+\beta A_{44} &=&0.  \nonumber
\end{eqnarray}
These equations imply that$\frac{A_{11}}{A_{44}}=\frac{\beta ^2}{\alpha ^2}$%
.With equation (23), we can delete the elements $A_{14}$ and $A_{41}$ from
equation (19) and (20). We rewrite $\rho $ and $\left| \rho ^T\right| ^T$ as
\begin{equation}
\rho =\left[
\begin{array}{llll}
A_{11}-\alpha ^2 & A_{12} & A_{31} & A_{32} \\
A_{21} & A_{22} & -\frac \alpha \beta (A_{11}+\beta ^2) & A_{42} \\
A_{13} & -\frac \alpha \beta (A_{11}+\beta ^2) & A_{33} & A_{34} \\
A_{23} & A_{24} & A_{43} & A_{44}-\beta ^2%
\end{array}
\right] ,  \eqnum{24}
\end{equation}
and
\begin{equation}
\left| \rho ^T\right| ^T=\left[
\begin{array}{llll}
A_{11}+\alpha ^2 & A_{12} & A_{31} & A_{32} \\
A_{21} & A_{22} & -\frac \alpha \beta (A_{11}-\beta ^2) & A_{42} \\
A_{13} & -\frac \alpha \beta (A_{11}-\beta ^2) & A_{33} & A_{34} \\
A_{23} & A_{24} & A_{43} & A_{44}+\beta ^2%
\end{array}
\right] .  \eqnum{25}
\end{equation}
Now we introduce Shur product (also called Hardmard product) \cite{horn} of $%
A=[a_{ij}]\in M_{m,n}$ and $B=[b_{ij}]\in M_{m,n}$. This Shur product is
given as $A\circ B=[a_{ij}b_{ij}]\in M_{m,n}$. It can be proved that $A\circ
B$ is positive semidefinite if $A$ and $B$ are positive semidefinite. Using
the definition of Shur product, we can get the relations between $\left|
\rho ^T\right| ^T$ and $\rho $. That is, $\left| \rho ^T\right| ^T=\rho
\circ S$ , where $S$ is defined as
\begin{equation}
S=\left[
\begin{array}{llll}
\frac{A_{11}+\alpha ^2}{A_{11}-\alpha ^2} & 1 & 1 & 1 \\
1 & 1 & \frac{A_{11}-\beta ^2}{A_{11}+\beta ^2} & 1 \\
1 & \frac{A_{11}-\beta ^2}{A_{11}+\beta ^2} & 1 & 1 \\
1 & 1 & 1 & \frac{A_{11}+\frac{\beta ^4}{\alpha ^2}}{A_{11}-\frac{\beta 4}{%
\alpha ^2}}%
\end{array}
\right]  \eqnum{26}
\end{equation}
we have used the relations between $A_{11}$\ and $A_{44}\ $to delete $A_{44}$%
. We have known that $\rho $ is positive semidefinite. Since Shur product
have the property that $A\circ B$ is positive semidefinite if $A$ and $B$
are positive semidefinite, we need only to consider when $S$ is positive
semidefinite under the condition $\rho \geq 0.$

At first, we need to get the relations of $A_{11},\alpha $ and $\beta $ by $%
\rho \geq 0$, that is
\begin{equation}
A_{11}\geq \alpha ^2,  \eqnum{27.1}
\end{equation}
\begin{equation}
A_{11}\geq \frac{\beta ^4}{\alpha ^2},  \eqnum{27.2}
\end{equation}
\begin{equation}
A_{22}A_{33}-\frac{\alpha ^2}{\beta ^2}(A_{11}+\beta ^2)^2\geq 0,
\eqnum{27.3}
\end{equation}
\begin{equation}
A_{11}-\alpha ^2+\frac{\alpha ^2}{\beta ^2}A_{11}-\beta ^2+A_{22}+A_{33}=1.
\eqnum{27.4}
\end{equation}
The last conditions is given by the condition $%
\mathop{\rm tr}%
(\rho )=1.$ Using conditions (27.3) to delete $A_{22}$ and $A_{33}$ in
(27.4), then the condition (27.4) can be rewritten as
\begin{equation}
A_{11}-\alpha ^2+\frac{\alpha ^2}{\beta ^2}A_{11}-\beta ^2+2.\frac \alpha %
\beta (A_{11}+\beta ^2)\leq 1.  \eqnum{28}
\end{equation}
That is
\begin{equation}
(\frac{\alpha +\beta }\beta )^2A_{11}\leq 1+\left( \alpha -\beta \right) ^2.
\eqnum{29}
\end{equation}
Without loss of generality, let $\alpha \geq \beta $, then the conditions of
$A_{11},\alpha $ and $\beta $ can be reduced to (27.1) and (29). In order to
make these two inequalities are consistent, it requires
\begin{equation}
\alpha ^2\leq \frac{1+\left( \alpha -\beta \right) ^2}{(\alpha +\beta )^2}%
\beta ^2  \eqnum{30}
\end{equation}
We can rewrite this condition in another way if let $k=\frac \alpha \beta $,
that is
\begin{equation}
\beta ^2\leq \frac 1{k^2(k+1)^2-(k-1)^2}.  \eqnum{31}
\end{equation}
In fact, this condition is about the negative eigenvalue of $\rho ^T$%
\begin{equation}
\left| E\right| =\alpha ^2+\beta ^2\leq E_1=\frac{1+k^2}{k^2(k+1)^2-(k-1)^2}.
\eqnum{32}
\end{equation}

Now we turn back to consider matrix $S$. Using conditions (27.1) and (27.2),
all of the $1\times 1$ and $2\times 2$ principal submatrices of $S$ are
positive semidefinite. We need only to consider $\det (S_{3\times 3})$ and $%
\det (S)$ where we choose a 3-by-3 principal submatrix as
\begin{equation}
S_{3\times 3}=\left[
\begin{array}{lll}
\frac{A_{11}+\alpha ^2}{A_{11}-\alpha ^2} & 1 & 1 \\
1 & 1 & \frac{A_{11}-\beta ^2}{A_{11}+\beta ^2} \\
1 & \frac{A_{11}-\beta ^2}{A_{11}+\beta ^2} & 1%
\end{array}
\right] .  \eqnum{33}
\end{equation}
Simple Calculation can show that

\begin{equation}
\det (S_{3\times 3})=\frac{4\nu [(2\mu -\nu )A_{11}+\mu \nu ]}{(A_{11}-\mu
)(A_{11}+\nu )^2},  \eqnum{34}
\end{equation}
where $\mu =\alpha ^2$ and $\nu =\beta ^{2\text{ }}$ for simplicity. It is
easy to verified that $\det (S_{3\times 3})\geq 0$ since we have let $\alpha
\geq \beta $ and $A_{11}\geq \alpha ^2$. It is also easy to get the
determinate of $S$ as
\begin{equation}
\det (S)=-\frac{8\nu [(\mu -\nu )^2A_{11}-2\mu \nu ^2]}{(A_{11}-\mu
)(A_{11}+\nu )^2(\mu A_{11}-\nu ^2)}.  \eqnum{35}
\end{equation}
For the same reason of $\det (S_{3\times 3}),$ to make $\det (S)\geq 0$, it
is sufficient to make $(\mu -\nu )^2A_{11}-2\mu \nu ^2\leq 0$. This
requirement is,
\begin{equation}
A_{11}\leq \frac{2\mu \nu ^2}{(\mu -\nu )^2}.  \eqnum{36}
\end{equation}
In order to make requirement (36) possible, $\alpha $ and $\beta $ must
satisfy the condition $\mu \leq \frac{2\mu \nu ^2}{(\mu -\nu )^2}$. That is,
\begin{equation}
1\leq \frac \mu \nu \leq \sqrt{2}+1  \eqnum{37}
\end{equation}
where we have considered the relation between $\alpha $ and $\beta $.

Now we have to compare the conditions (29) and (36). If
\begin{equation}
\frac{1+\left( \alpha -\beta \right) ^{2}}{(\alpha +\beta )^{2}}\beta
^{2}\leq \frac{2\alpha ^{2}\beta ^{4}}{(\alpha ^{2}-\beta ^{2})^{2}},
\eqnum{38}
\end{equation}%
then the requirement (36) will be automatically guaranteed by the conditions
(27.1) and (29) (positive semidefinite of $\rho $). That is, $S$ will be
positive semidefinite for $\rho \geq 0.$ This condition is also about the
negative eigenvalue of $\rho ^{T}$%
\begin{equation}
\left\vert E\right\vert =\mu +\nu \geq E_{2}=\frac{(1+k^{2})(k-1)^{2}}{%
2k^{2}-(k-1)^{4}}.  \eqnum{39}
\end{equation}%
So we have two conditions about the negative eigenvalue, conditions (32) and
(39), coming from the positive semidefinite of $\rho $ and $S$,
respectively. In order to make these two conditions consistent, it will
require that
\begin{equation}
\frac{1}{k^{2}(k+1)^{2}-(k-1)^{2}}\geq \frac{(k-1)^{2}}{2k^{2}-(k-1)^{4}}.
\eqnum{40}
\end{equation}%
this inequality gives the same constraint on the parameter $k$ as condition
(37). In another word, inequality (40) is automatically satisfied under the
condition (37).

So when the eigenvector corresponding to the negative eigenvalue of $\rho ^T$
satisfies condition $1\leq k\leq \sqrt{\sqrt{2}+1}$ which implies the
condition of negative eigenvalue about $E_2\leq \left| E\right| \leq E_1$,
the formula $\left| \rho ^T\right| ^T=\rho \circ S$ is positive semidefinite.

This is the end of the proof.

Q. E. D.

When the density matrix $\rho $ is separable, this theorem is trivial. Our
theorem shows that the entanglement of the eigenvector corresponding to
negative eigenvalue of $\rho ^{T}$ is concerned with the formula (1). When
the entanglement is zero or sufficiently high, this formula is true. As a
matter of fact, the conditions (17.1) and (17.2) are concerned with the
entanglement of $\left\vert \Psi \right\rangle $. Equation (17.1) means that
the state is a product state and the entanglement is zero. Equation (17.2)
means that the state is near the maximal entangled state where $\alpha
=\beta $ and the entanglement is rather high. Our theorem also shows that
this formula is related with the negative entropy of the density matrix $%
\rho $. We know that the negativity\cite{N} of a state $\rho $ is defined as
$N(\rho )=(\left\vert \rho ^{T}\right\vert _{1}-1)/2$ (another measurement,
logarithmic negativity, is defined as E$_{N}=\log _{2}\left\vert \rho
^{T}\right\vert _{1}$). It is easy to show $N(\rho )=\sqrt{\left\vert
E\right\vert }$. So it is clear that the condition of equation (40) is just
for the negativity of state $\rho $. In our case, the negativity is bounded
by the formula of $k$ (in another word, the entanglement of $\left\vert \Psi
\right\rangle $). So the negativity of $\rho $ and the entanglement of $%
\left\vert \Psi \right\rangle $ are connected. We cannot determine which
aspect is the key quantity concerned with the formula (1).

\section{Summary}

In this paper, we discussed the mathematical property of partial
transposition which is a famous operator in quantum information. More
concretely, we consider the number of the negative eigenvalues of $\rho ^{T}$%
. Through Monte Carlo random method, we find that there are good reasons to
conjecture that the maximal number of negative eigenvalues of $\rho ^{T}$
(where $\rho $ is a quantum state in $C^{N\times N}$ Hilbert space) is $%
\frac{N(N-1)}{2}$. This conjecture is clear, but it is hard to prove even
for the two-qubit system. We just prove it under some special case, but this
special case have included many often used cases. How to completely prove
this result is still a big challenge for us. There are some reasons for its
difficulty. One is that the PT transformation has no direct relations with
some known geometrical or algebraic operators. So we can't use their
powerful tools. The other is that the eigenvalues of $\rho ^{T}$ are
invariants under global unitary in the whole Hilbert space. But $\rho $ and $%
\rho ^{T}$ are related by their local elements, these relations are not
invariants under global unitary.

When the state $\rho $ is separable, it is easy to see that $\left\vert \rho
^{T}\right\vert ^{T}\geq 0$ since the operator $\left\vert .\right\vert $
does nothing on $\rho ^{T}$. We have shown that it is also true when the
entanglement of the eigenvector corresponding to the negative eigenvalue of $%
\rho ^{T}$ is sufficiently high (or the negative entropy of $\rho $ satisfy
some condition). But to complete this proof is beyond the technique
introduced in this paper. $\left\vert .\right\vert $ is only concerned with
the eigenvalues of $\rho ^{T},$ but PT transformation is concerned with the
elements of $\rho $. So it is hard to find the direct relations between $%
\rho $ and $\left\vert \rho ^{T}\right\vert ^{T}$. In other words, it is
hard to prove the positive semidefinite of $\left\vert \rho ^{T}\right\vert
^{T}$ by the positive semidefinite of $\rho $. There are some evidences that
we need consider all of the elements of $\rho $ to complete the proof. But
it is hard to deal with so many variables. So it may be necessary to
introduce some new techniques to this problem.

\section{Acknowledgment}

This work was funded by the National Fundamental Research Program
(2001CB309300 ), the Innovation Funds from Chinese Academy of Sciences
(CAS), China Postdoctoral Science Foundation(2005038012) and Chinese Academy
of Sciences K. C. Wong Post-doctoral Fellowships.

\end{document}